\documentclass[aps,prb,amsfonts,amssymb,twocolumn,amsmath,floatfix,showpacs]{revtex4-1}
\usepackage{amsbsy}
\usepackage[dvips]{color}
\usepackage[dvips]{graphics}
\usepackage{graphicx}
\usepackage{bm}

\DeclareMathOperator{\sgn}{sgn}

\begin{document}

\title{Anharmonic Josephson current in junctions with an
interface pair breaking}

\author{Yu.\,S.~Barash}
\affiliation{Institute of Solid State Physics, Russian Academy of Sciences,
Chernogolovka, Moscow District, 142432 Russia}

\date{\today}

\begin{abstract}
Planar superconducting junctions with a large effective Josephson
coupling constant and a pronounced interface pair breaking are shown
to represent weak links with small critical currents
and strongly anharmonic current-phase relations. The supercurrent
near $T_c$ is described taking into account the interface pair
breaking as well as the current depairing and the Josephson
coupling-induced pair breaking of arbitrary strengths. A new
analytical expression for the anharmonic supercurrent, which is in
excellent agreement with the numerical data presented, is obtained.
In junctions with a large effective Josephson coupling constant and a
pronounced interface pair breaking, the current-induced depairing is
substantially enhanced in the vicinity of the interface thus having a
crucial influence on the current-phase relation despite a small
depairing in the bulk.
\end{abstract}


\maketitle

The Josephson current is one of the remarkable manifestations of
quantum coherence on the macroscopic scale in condensed matter
physics. The supercurrent depends on the phase difference of
the order parameters across the junction interface.
The study of the current-phase relation (CPR) in the junctions
makes it possible to identify physical processes, which form
supercurrents under diverse conditions. It is also beneficial for
junction applications. The problem attracted much attention while
studying both highly transparent junctions with strongly anharmonic
CPRs and tunnel junctions, where the second harmonic of the
supercurrent comes into play due to the suppression of the first
one~\cite{Golubov2004,tsueikirt00,Buzdin2005,Geshkenbein1986,%
Tanaka1994,Yip1995,Il'ichev1999,Mannhart2002,Lindstrom2003,%
Lindstrom2006}. The latter takes place in the junctions involving
unconventional superconductors with special interface-to-crystal
orientations and at $0$-$\pi$ transitions.

One of the earlier theoretical results, clarifying a variety of
aspects of the problem, is the anharmonic CPR for the superconducting
point contacts~\cite{Kulik1975,Kulik1977,Habekorn1978,Likharev1979,
Zaitsev1984}. Due to the negligibly small current-induced pair
breaking at any transparency value, the theory of point contacts
is simplified. The depairing plays an important role in forming
anharmonic CPRs in highly transparent planar junctions, unlike its
negligible role in point contacts. Since the critical current $j_c$
of usual planar junctions becomes, with increasing
transparency, comparable with the depairing current $j_{dp}$ in the
bulk, the junctions do not represent weak links. In other words, in
the junctions, the current-induced depairing brings about a pronounced
anharmonicity only in the crossover from the tunnel Josephson current
to the bulk superflow \cite{Kupriyanov1992,Sols1994,Freericks2002,%
Golubov2004}.

There are at least three types of pair breaking processes taking
place in charge transport in the superconducting junctions:
the pair breaking produced by the phase-dependent Josephson coupling,
by the current and by the interface itself. These are the very same
effects which can lead to a noticeably anharmonic CPR not only at low
or intermediate temperatures, but also near $T_c$.
Here, I show that planar junctions with a large effective Josephson
coupling constant and a pronounced interface pair breaking can
possess strongly anharmonic CPRs and small critical currents
satisfying the condition $j_c\ll j_{dp}$. An enhancement of the
current-induced depairing near the interface will be identified.
The anharmonic supercurrent
near $T_c$ will be obtained within the Ginzburg-Landau (GL) theory in
the presence of all three types of the pair breaking processes of
arbitrary strengths. Along with the numerical solution based on GL
equations, a new analytical CPR will be derived and shown to be in
excellent agreement with the numerical data in a wide range of
parameters. For tunnel junctions the obtained results present a new
description of higher harmonics of the supercurrent and extend the
known expressions for the first and second harmonics to include the
effects of interface pair breaking.

The CPRs obtained earlier near $T_c$ with the microscopic boundary
conditions for standard dirty $s$-wave junctions
\cite{Ivanov1981,Kupriyanov1992}, have been considered in literature
solely as the particular properties of the specific systems
\cite{Golubov2004}. The anharmonic CPR obtained in this paper, and
influenced by the interface pair breaking, is of general form
inherent in the GL theory, and is applicable to a variety of planar
junctions including those containing $d_{x^2-y^2}$-wave superconductors
and/or magnetic interlayers.

The free energy functional for Josephson junctions near $T_c$ results
in the GL equations and the boundary conditions (BC) for them
\cite{deGennes1966,Yip1990,Sigrist1991,Lifshitz1995, Mineev1999}.
Consider symmetric junctions with a spatially constant width, which
is much less than the Josephson penetration length, and with a plane
interlayer at $x=0$ of zero length within the GL approach.  Assume
the usual form of the GL free energy, which applies, for example, to
$s$-wave and $d_{x^2-y^2}$-wave junctions.  If the Josephson coupling
$g_J|\Psi_+ -\Psi_-|^2$ is strong, not only this term but all the
interface and bulk contributions to the free energy generally
participate in the formation of CPRs as a consequence of the
dependence of absolute values of the order parameters at the
interface on the phase difference. This concerns, in particular, the
gradient bulk term $K|\nabla\Psi |^2$ and the interface contribution
of the form $g(\left| \Psi_+ \right|^2+\left| \Psi_-\right|^2)$.

Moving on to the order parameter $f(x)e^{i\chi(x)}$ normalized to
$f=1$ in the bulk without superflow, one gets the first integral of
the GL equation in the presence of the supercurrent
\cite{Langer1967} in the form of
\begin{equation}
\left(\frac{df}{d\tilde{x}}\right)^2+f^2\!-
\frac{1}{2}f^4\!+\,\dfrac{4\tilde{j}^2}{27f^2}=
2f_{\infty}^2-\dfrac{3}{2}f_{\infty}^4.
\label{gl1d8}
\end{equation}
Here $\tilde{x}=x/\xi$, $\xi=\xi(T)$ is the superconducting coherence
length, $\tilde{j}$ is the spatially constant normalized current
density $\tilde{j}={j}/{j_{dp}}=-(3\sqrt{3}/{2})({d\chi}/{d\tilde{x}}
)f^2$ and $f_{\infty}$ is the asymptotic value of $f$ in the depth of
the superconducting leads.

The BC introduce in the GL theory at least two characteristic lengths
$\ell=K/g_J$, $\delta=K/g$. The effective dimensionless Josephson
$g_\ell=g_J\xi(T)/K$ and interface $g_{\delta}=g\xi(T)/K$ coupling
constants, associated with these lengths, will be used below. For
symmetric junctions with $f$ continuous through the interface, the BC
for $f$ as well as the expression for the Josephson current via
$f_{0}$ and the phase difference $\chi=\chi_--\chi_+$ at the
interface, are obtained from the BC for complex order parameters:
\begin{equation}
\left(\!\frac{df}{d\tilde{x}}\!\right)_{\pm}\!\!\!=
\pm\!\left(g_{\delta}+2g_\ell\sin^2\!\frac{\chi}2\right)\!f_0,
\enspace
\tilde{j}=\frac{3\sqrt{3}}{2}g_\ell f_{0}^2\sin\chi.
\label{bcss99}
\end{equation}
Here the effective phase-dependent extrapolation length $b(\chi)=
\left({\delta}^{-1}+2{\ell}^{-1}\sin^2(\frac{\chi}2)\right)^{-1}$
controls the pair breaking produced by the phase difference and by
the interface. Let's denote $g_b(\chi)=(g_{\delta}+2g_\ell
\sin^2\frac{\chi}2)$.

Since the material parameters in the normal state ${g}_J$ and ${g}$
are not assumed to depend here on $T$ near $T_c$, one should have
$|g_{\ell}|\gg 1$ and/or $|g_{\delta}|\gg 1$ quite close to $T_c$ due
to large values of $\xi(T)$. However, the coupling constants $g_J$
and $g$ can themselves be very small and the temperature range with
large $g_\ell$ and/or $g_{\delta}$ be too narrow, as it occurs in
standard tunnel junctions. Due to a very small surface pair breaking in
conventional s-wave junctions, one parameter $g_\ell$ is usually
assumed to describe the interfaces in \eqref{bcss99} rather than both
$g_{\ell}$ and $g_{\delta}$ as is in the regular case. At $\chi =0$, such
symmetric junctions contain no pair breaking at all, and the BC
\eqref{bcss99} is reduced to $({df(0)}/{dx})=0$.

If $g_J$ and/or $g$ were very small, one would need to introduce into
\eqref{bcss99} the terms of the next order of smallness, in
particular, in powers of the order parameter. Such terms could be of
importance and bring about additional phase dependence and
material-dependent parameters to the problem. Here, only the simplest
conditions will be assumed, when \eqref{bcss99} applies to a wide
range of values of $g_\ell$ and $g_{\delta}$. This agrees with the
microscopic model results \cite{deGennes1966,Galaiko1969,Bratus1977,%
Ivanov1981,Svidzinskii1982,Geshkenbein1988,Kupriyanov1992} and, for
instance, takes place within the GL approach for sufficiently large
values of $g_{\ell}$ and $g_{\delta}$, which is the particular focus
of this paper.

There is no need to solve differential equation \eqref{gl1d8}
in order to find $f_{0}$, and, consequently, to find $\tilde{j}$
via \eqref{bcss99}. One puts $x=0$ in \eqref{gl1d8} and, using
\eqref{bcss99}, eliminates the current and the first derivative of
the order parameter. This results in a biquadratic relation between
the self-consistent values of $f^2_{0}$ and $f_{\infty}^2$. The second
relation between them follows from the current conservation and the
asymptotic formulas in the bulk. The current-induced depairing in the
bulk is conveniently described via the superfluid velocity
$\tilde{j}=(3\sqrt{3}/2)\tilde{v}_s(1-\tilde{v}_s^2)$,\,\,$
f_{\infty}^2=1-\tilde{v}_s^2$ \cite{Bardeen1962, Tinkham1996}.
Equating the asymptotic expression for the current to that in
\eqref{bcss99} with $f_{0}^2=(1-\tilde{v}_s^2)\alpha$ one obtains
$\tilde{v}_s=\alpha g_\ell\sin\chi$. Considering that both quantities
$f_0$ and $f_{\infty}$ as well as the current itself are now expressed
via the only variable $\alpha$, the fourth-order polinomial equation for
$\alpha$ follows from the biquadratic relation between $f_0$ and
$f_{\infty}$
\begin{equation}
2g_b^{2}(\chi)\alpha-
(1-\alpha)^2[1-\alpha(\alpha+2)g_\ell^{2}\sin^2\chi]=0.
\label{alphaeq}
\end{equation}
Eq.~\eqref{alphaeq} is exact within the conventional GL approach with
BC \eqref{bcss99}. In the particular case of standard $s$-wave
junctions, $g_b(\chi)=2 g_\ell\sin^2(\chi/2)$. Then \eqref{alphaeq} is
reduced to Eq.(8) of Ref.~\cite{Kupriyanov1992}, if one corrects a
misprint $\Gamma_B\to\Gamma_B^2$ in (8) and identifies the parameter
of the GL theory $g_\ell^{-1}=\ell/\xi$ with the model parameter
$\Gamma_B$ entering the microscopic BC for dirty $s$-wave
superconductors.

An analytical solution of the problem can be obtained
assuming a small depairing in the bulk $\tilde{j}^2 \ll 1$ that
allows to use $f_{\infty}^2\approx1-(4/27) \tilde{j}^2$ and to
disregard the smaller terms on the right-hand side of \eqref{gl1d8}.
Then one gets from \eqref{gl1d8}, \eqref{bcss99} a biquadratic
equation for $f_{0}$, which results in the analytical solution for
the CPR:
\begin{widetext}
\begin{equation}
\tilde{j}\left(g_{\ell},g_{\delta},\chi\right)=
\frac{3\sqrt{3}g_{\ell}\sin\chi}{2(1+2g_{\ell}^{2}
\sin^2\chi)}\biggl[1+g_b^{2}(\chi)+
g_{\ell}^{2}\sin^2\chi
-\sqrt{\bigl(g_b^{2}(\chi)+g_{\ell}^{2}\sin^2\chi\bigr)^2+
2g_b^{2}(\chi)}\,\biggr].
\label{bcss1012p}
\end{equation}
\end{widetext}
Since only higher order terms begining with $\propto \tilde{j}^{4}$
have been neglected in its derivation, the CPR \eqref{bcss1012p}
turns out to describe the current behavior almost perfectly, if
$\tilde{j}< 0.7$. For $\tilde{j}>0.7$ it gives a good interpolation
of the numerical solution based on \eqref{alphaeq}, resulting in
the deviations not exceeding $10\%$.

As seen in \eqref{alphaeq} and \eqref{bcss1012p}, the anharmonic
Josephson current $\tilde{j}$ depends, in general, on the two
dimensionless effective coupling constants $g_\ell$, $g_{\delta}$ and
the phase difference $\chi$. According to the simple physical
arguments as well as the microscopic results \cite{Galaiko1969,%
Bratus1977,Svidzinskii1982,Geshkenbein1988}, a variation of tunneling
parameters principally modifies $g_{\ell}$, while the surface pair
breaking mostly contributes to $g_{\delta}$. This signifies that the
junction transparency $D$ enters the combination of microscopic
parameters representing $g_{\ell}$. The last statement agrees with
the microscopic results for $s$-wave junctions with nonmagnetic
interfaces~\cite{Galaiko1969,Bratus1977,Ivanov1981,Svidzinskii1982,%
Kupriyanov1992}, where the corresponding combination is sometimes
identified as the effective transparency \cite{Bezuglyi1999,Bezuglyi2005}.
The microscopic estimations of the effective Josephson coupling constant
$g_\ell$ directly follow from those results. In the $s$-wave tunnel
junctions ($D \ll1$), one gets $g_\ell\sim D\xi(T)(l^{-1}+\xi_0^{-1})
$, where $l$ is the mean free path. In dirty superconductors, the
ratio $\xi(T)/l$ can easily reach $100$ even at low temperatures.
Hence, for small and moderate transparencies, the quantity $g_\ell\sim
D\xi(T)/l$ can vary from vanishingly small values in the tunneling
limit to those well exceeding $100$ near $T_c$. In highly transparent
junctions ($(1-D)\ll 1$) the parameter $g_{\ell}\propto (1-D)^{-1}$
can be arbitrary large \protect\footnote{The solutions based on
\eqref{alphaeq} or \eqref{bcss1012p} satisfy the relation
$|g_b| f_0\alt 1$, in particular, at large values of $|g_\ell|$
and/or $|g_\delta|$. In view of \eqref{bcss99}, this agrees with
the condition that a strong suppression of the order parameter on
each side of the interface takes place on a scale comparable with
$\xi(T)$.}. The quantity $g_\ell$ can also take on negative values,
which correspond to $\pi$-junctions, as seen in \eqref{bcss1012p}.

The range of variation of the interface coupling $g_\delta$ can
likewise be quite wide. For $s$-wave superconductor-insulator
interfaces, the Josephson coupling vanishes and the extrapolation
length $b$ is reduced to $\delta$. The microscopic estimations of
$\delta$ in such cases show it to be very large usually resulting in
a negligibly small contribution to the BC, unlike the
superconductor-normal metal interfaces \cite{deGennes1966}.  The
length $\delta$ can vary widely for $d$-wave superconductor-insulator
flat surfaces, where it substantially depends on surface-to-crystal
orientations \cite{Barash1995,Alber1996, Agterberg1997}. Although in
this case $\delta$ is strongly influenced by the surface roughness,
in particular, by faceting \cite{Mannhart2002}.

\begin{figure}[t]
\begin{minipage}{.49\columnwidth}
\includegraphics[width=.99\columnwidth,clip=true]{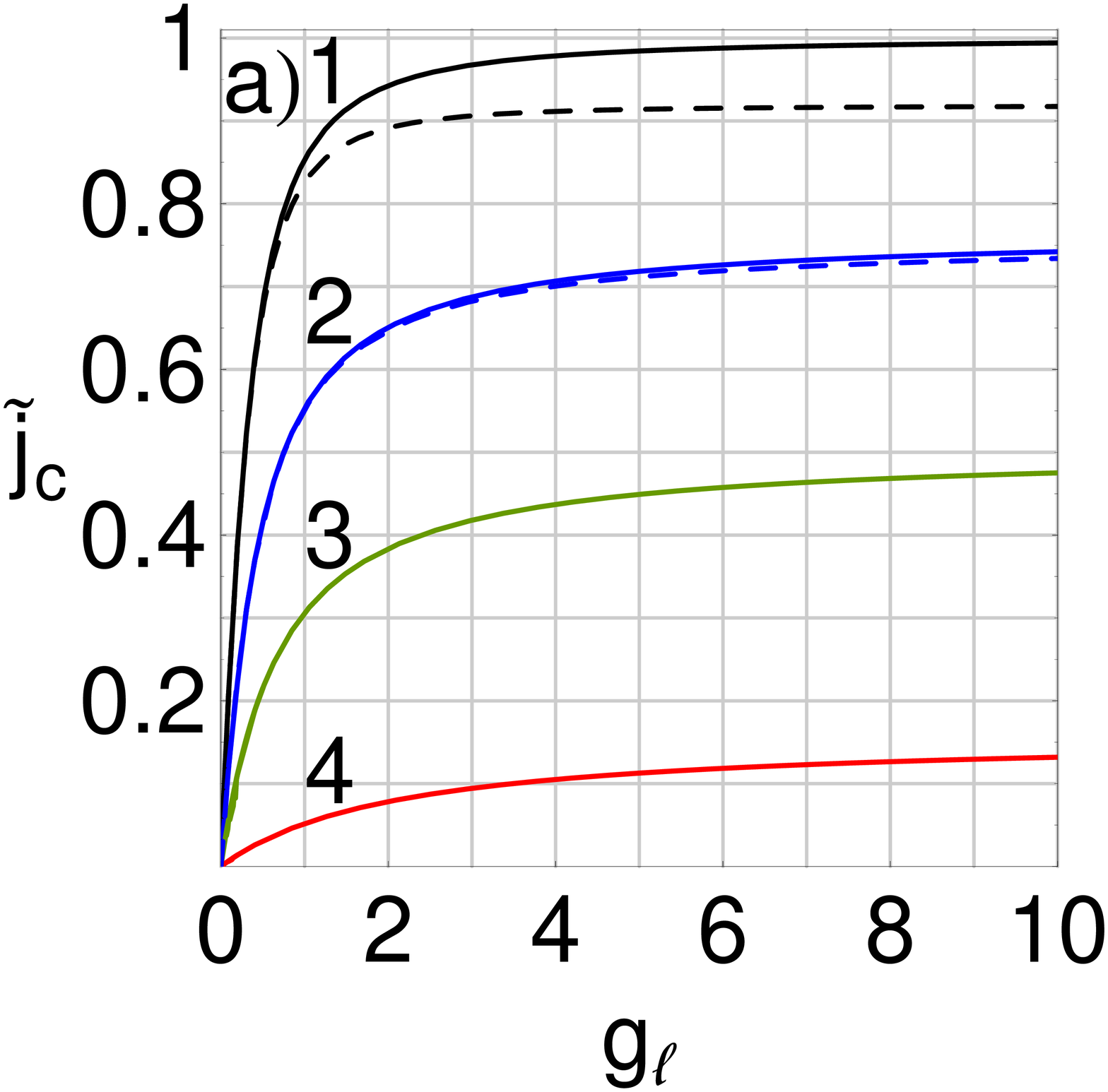}
\end{minipage}
\begin{minipage}{.49\columnwidth}
\includegraphics[width=.98\columnwidth,clip=true]{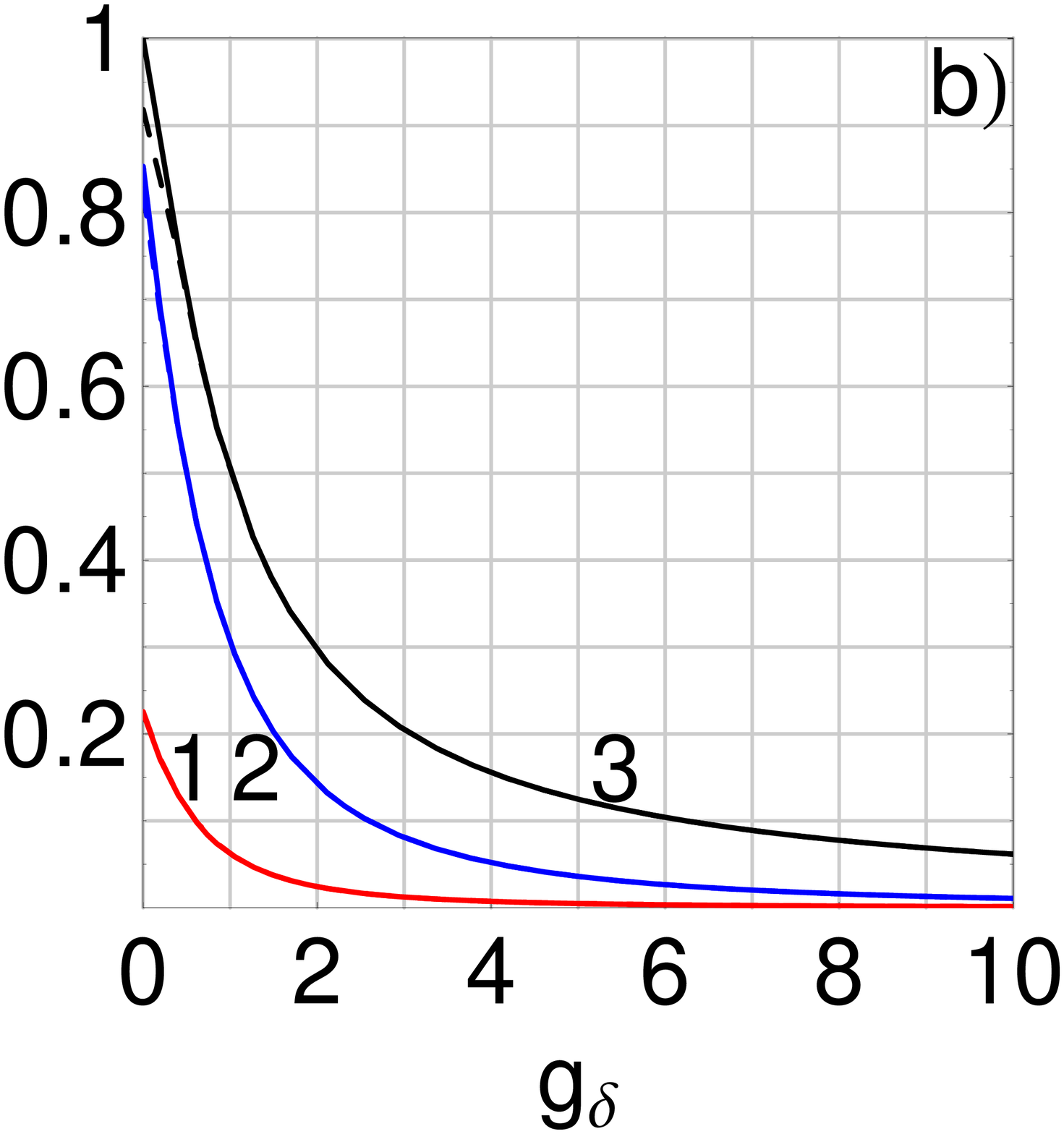}
\end{minipage}
\caption{ a) $\tilde{j}_c$ as a function of $g_\ell$, taken for various
$g_{\delta}$:\, 1.\,$g_{\delta}\!=\!0$,\,\,2.\,$g_{\delta}\!=\!0.4$,\,\,%
3.\,$g_{\delta}\!=\!1$,\,\,4.\,$g_{\delta}\!=\!4$.\,%
b) $\tilde{j}_c$ as a function of $g_{\delta}$, taken for various
$g_\ell$:\, 1.\,$g_\ell=0.1$,\,\, 2.\,$g_\ell=1$,\,\, 3.\,$g_\ell=100$.
\label{anharmon1ab} }
\vspace{-0.2cm}
\end{figure}

A regular situation is characterized by a local suppression of the
order parameter at the interface. For this condition to hold, the
effective extrapolation length $b(\chi)$ should be positive at any
phase difference and, hence, $g_{\delta}, (g_{\delta}+2g_\ell)>0$.  A
superconducting state occurs locally near the interface above the
bulk $T_c$ under the opposite condition $b(\chi)<0$ with $\chi$
ensuring the free energy minimum \cite{Andreev1987}. Only the
simplest conditions $g_{\delta,\ell}>0$ will be analyzed in detail in
this paper, although the main results obtained here apply to
substantially more general circumstances. Other conditions, including
magnetic field effects and/or negative $g_{\delta,\ell}$, will be
studied elsewhere.

Figs.~\ref{anharmon1ab}\,a, b show the critical current $\tilde{j}_c$
as a function of coupling constants $g_\ell$ and $g_\delta$. Solid
curves have been calculated based on \eqref{alphaeq}. Dashed curves
correspond to the analytical expression \eqref{bcss1012p}. Only for a
small interface pair breaking ($g_{\delta}\alt 1$) and for
$g_\ell\agt 1$, the current $\tilde{j}_c$ becomes comparable with
$1$, i.e., with the deparing current in the bulk. Thus the condition
$g_\ell\gtrsim 1$ is the hallmark of a strong Josephson coupling.
Comparatively small deviations of dashed curves from the solid ones
are discernible only when the current exceeds about 0.7. With
increasing $g_{\delta}$, the growing interface pair breaking
suppresses the critical current. For $g_{\delta} \gtrsim 4$, the
critical current remains quite small $\tilde{j}_c\ll 1$ at any
$g_{\ell}$, which would normally occur in conventional tunnel
junctions with small effective transparencies. In other words, in the
regime of strong interface pair breaking $g_\delta>4$, the junctions
represent weak links at any $g_{\ell}$, including $g_{\ell}\agt 1$.

Though \eqref{bcss1012p} is a combined result of all depairing
effects, the origin of its characteristic anharmonic features is
traced back unambiguously. The whole of the phase-dependence in
\eqref{bcss1012p}, except for that contained in $g_b(\chi)$, is
generated by the current via $f_{\infty}$ on the right hand side or
by the last term on the left hand side in \eqref{gl1d8}. Such
dependence would retain the CPR \eqref{bcss1012p} unchanged under the
transformation $\chi\to\pi-\chi$. The symmetry is destroyed by the
phase dependence of $g_b^2(\chi)$, which originates from the BC
\eqref{bcss99} and can become pronounced, if $ |g_{\delta}|\alt
2|g_{\ell}|$.

Whereas the CPR \eqref{bcss1012p} is derived by assuming small
depairing effects in the bulk, the depairing can be of crucial
importance in \eqref{bcss1012p} within its domain of applicability.
This is the case in the presence of a pronounced interface pair
breaking, where an enhancement of the current-induced depairing,
unlike the bulk, occurs near interfaces of junctions with $g_\ell
\gg1$. In particular, the phase-dependent term in the denominator in
\eqref{bcss1012p}, which is directly induced by the depairing, plays
a key role in the case $g_\ell\gg1$ in restricting the normalized
current value. The bracketed expression in the denominator originates
from the coefficient before $f_0^4$ in the biquadratic equation for
$f_0$. The relative depairing correction coming from the bulk is
$(8/27)\tilde{j}^2=2g_\ell^{2}f_0^4\sin^2\chi $ and its smallness
signifies $2g_\ell^{2}\sin^2\chi f_0^4\ll1$. As seen, the term
$2g_\ell^{2}\sin^2 \chi$ in the denominator is
allowed to exceed the unit considerably, when the condition
$2g_\ell^{2}\sin^2\chi f_0^4\ll1$ holds at the expense of a strongly
suppressed order parameter at the interface $f_0^4\ll1$.  Numerical
results corroborate that, if $g_\delta\agt4$, the condition is
satisfied at any $g_\ell$ including $g_\ell\gg1$ (see also
Figs.~\ref{anharmon1ab}\,a, b). This validates keeping
\eqref{bcss1012p} without its expanding in powers of
$g_\ell^{2}\sin^2\chi$ and explains the quantitative applicability of
\eqref{bcss1012p} to junctions with  the pronounced interface pair
breaking at arbitrary $g_\ell$.

A number of specific CPRs follow from \eqref{bcss1012p} under a
variety of particular conditions. Consider here two basic examples.
The tunneling limit shows up in \eqref{bcss1012p} under the condition
$|g_\ell|\ll1$. Developing \eqref{bcss1012p} as series in $g_\ell$ at
any value of $g_{\delta}$, one obtains numerous harmonics whose
weight is determined by $g_\ell$ and $g_{\delta}$ rather than by the
transparency itself. The first and the second order terms result in
\begin{equation}
\tilde{j}\!\approx\!\tilde{j}_{c1}^{(1)}\!\!\left[\sin\chi
-\frac{2g_{\ell}\sgn(g_{\delta})}{\sqrt{2+g_{\delta}^2}}\!\left(\!\sin\chi
-\frac12\sin2\chi\!\right)\right]\!.
\label{gl1d21}
\end{equation}
Here $\tilde{j}_{c1}^{(1)}=({3\sqrt{3}}/4)g_\ell\left(\sqrt{2+g_{
\delta}^2}-\left|g_{\delta}\right|\right)^2$ is the main contribution
to the first harmonic $\tilde{j}_1=\tilde{j}_{c1}\sin\chi$ that is
applicable at any $g_{\delta}$. Under the
condition $|g_{\delta}|\ll 1$ it is reduced to the
well-known result for tunnel junctions
$\tilde{j}_{c0}\equiv(3\sqrt{3}/2)g_\ell$, which is only justified
when disregarding the interface pair breaking. In the opposite limit
$g_{\delta}^2 \gg 1$ the pair breaking strongly suppresses the
current and $\tilde{j}_{c1}^{(1)}\approx\tilde{j}_{c0}/
(2g_{\delta}^2)\ll\tilde{j}_{c0}$, as is also known
\cite{deGennes1966,Geshkenbein1986,Deutscher1987,Yip1990,Sigrist1991,%
Barash1995,Mineev1999}. In particular, the original current $j_{c1}^{(1)}=
\tilde{j}_{c1}^{(1)}j_{dp}\propto (T_c-T)$ for $|g_{\delta}|\ll1$ and
$j_{c1}^{(1)} \propto(T_c-T)^2$ for $g_{\delta}^2 \gg 1$ near $T_c$.
The second order terms in $g_\ell$ bring about the main contribution
to the second harmonic $\tilde{j}_2= \tilde{j}_{c2}\sin2 \chi$ as
well as corrections to the first one.  The relative weight of the
second harmonic in \eqref{gl1d21} diminishes with increasing $
g_\delta^2$. The sign of $\tilde{j}_{c1}$ coincides with the sign of
$g_\ell$, while the sign of $j_{c2}$ is determined by the sign of
$g_\delta$. For small pair breaking $0<g_\delta\ll 1$ the second
order term $ \propto g_\ell^2$ is simplified to the following
correction to the current $-\sqrt{2}j_{c0}g_{\ell}(\sin\chi-(1/2)
\sin2\chi)$, in agreement with the corresponding microscopic
results~\cite{Antsygina1973,Kupriyanov1992} for dirty and pure
$s$-wave junctions. Note that the phase dependence generated by the
current depairing shows up in \eqref{bcss1012p} beginning with the
third order terms in $g_{\ell}$.

\begin{figure}[t]
\begin{minipage}{.49\columnwidth}
\includegraphics[width=.99\columnwidth,clip=true]{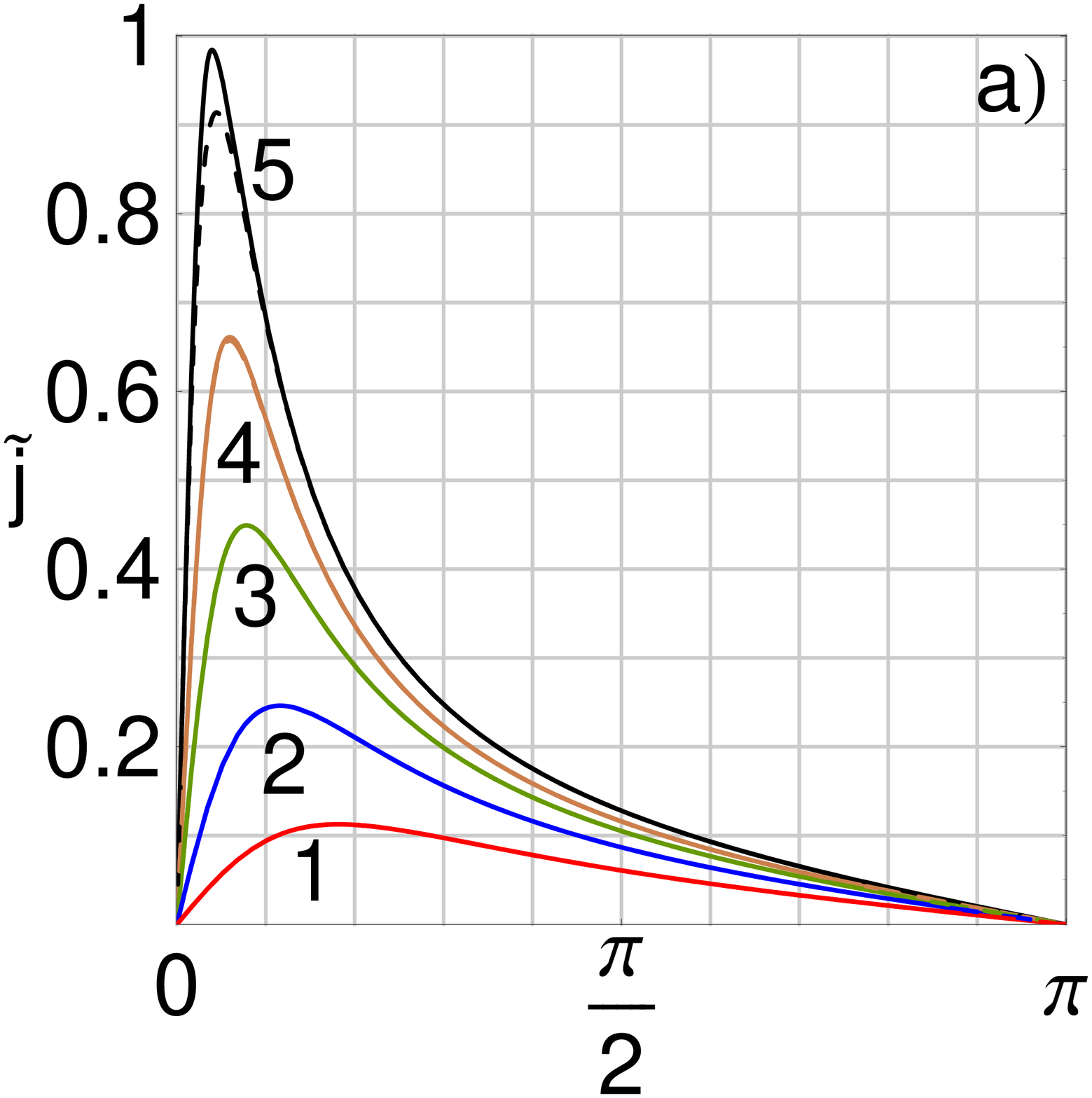}
\end{minipage}
\begin{minipage}{.49\columnwidth}
\includegraphics[width=.99\columnwidth,clip=true]{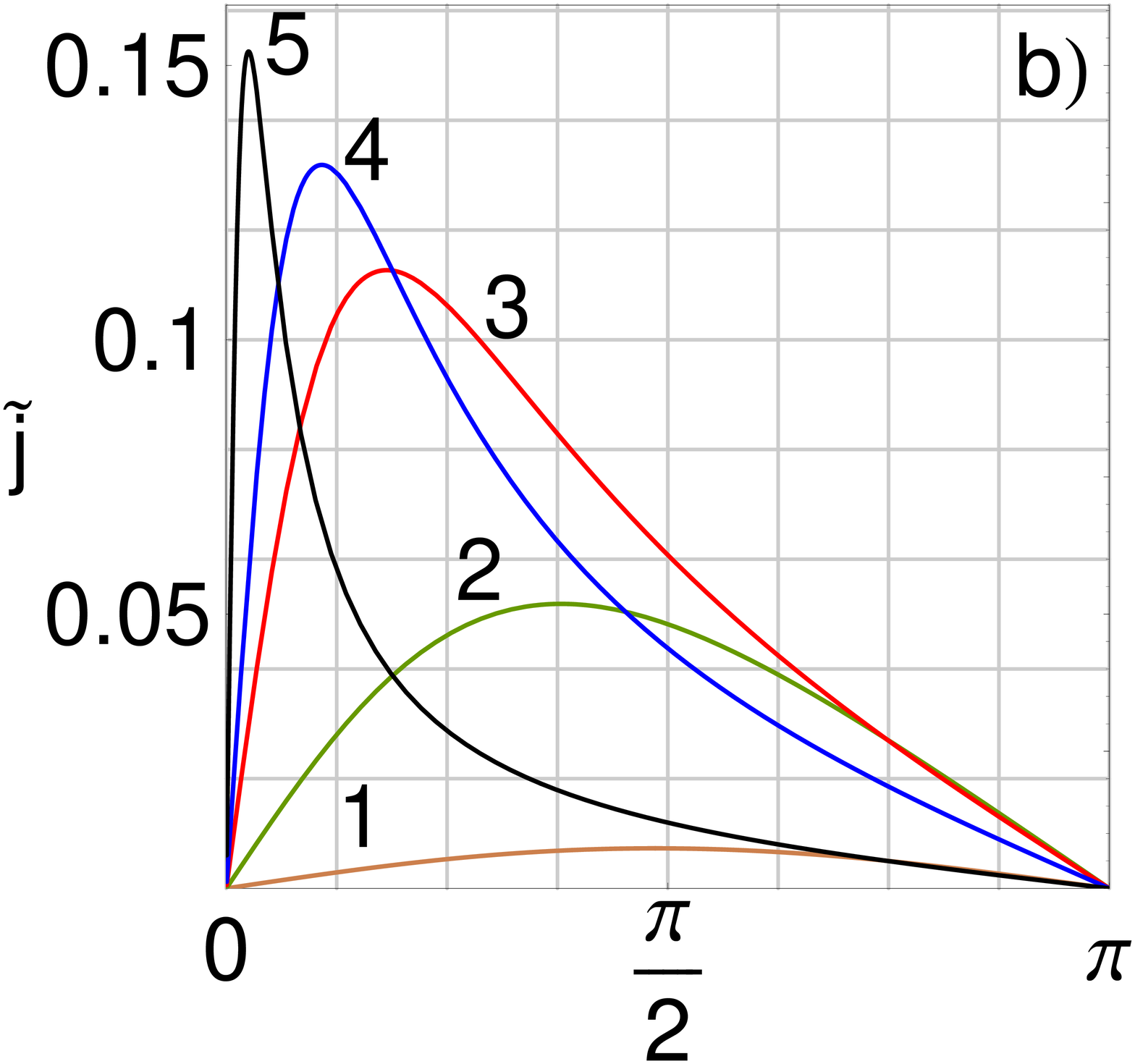}
\end{minipage}
\caption{a) CPRs $\tilde{j}(\chi)$
for $g_\ell=\!5$ and various $g_{\delta}$:\, 1.\,$g_{\delta}=4$,\,\,%
2.\,$g_{\delta}=2$,\,\,3.\,$g_{\delta}=1$,\,\,4.\,$g_{\delta}=0.5$\,\,\,%
5.\,$g_{\delta}=0$.\,
b) CPRs for $g_{\delta}=4$ taken for various $g_\ell$:\, 1.\,%
$g_\ell=0.1$,\,\,2.\,$g_\ell=1$,\,\,3.\,$g_\ell=5$,\,\,4.\,$g_\ell=10$,\,\,%
5.\,$g_\ell=50$.
\label{anharmon2ab} }
\vspace{-0.2cm}
\end{figure}

The second example reveals the strongly anharmonic features contained in
\eqref{bcss1012p}. Consider junctions with the strong interface pair
breaking $g_{\delta}^2\gg 1$. Then a comparatively simple
approximate expression follows from~\eqref{bcss1012p}
\begin{equation}
\tilde{j}\approx\frac{3\sqrt{3}g_\ell\sin\chi}{4[
g_\delta^2+4(g_\delta+g_\ell)g_\ell
\sin^2\frac{\chi}2]}.
\label{cpr3}
\end{equation}
The corresponding critical current $\tilde{j}_c=3\sqrt{3}g_\ell/
4|g_\delta(g_\delta+2g_\ell)|\ll 1$ is always small. The associated
phase difference is determined by the relation $\sin\chi_c=|g_\delta
(g_\delta+2g_\ell)|/[(g_\delta+g_\ell)^2+g_\ell^2]$. It varies widely:
$\chi_c$ is small $\approx(g_{\delta}/g_\ell)$, if $g_\ell\gg g_\delta$,
and approaches $\pi/2$ in the opposite limit $g_\delta\gg g_\ell$.
Strongly anharmonic CPRs show up in \eqref{cpr3} under the conditions
$g_\ell^2\gg g_\delta^2\gg 1$. Also
one has $j_c\propto (T_c-T)^2$. Thus, at finite $g$, the temperature dependence
$j_c(T)$ is quadratic quite close to $T_c$, where $g_{\delta}\gg1$. With
increasing
$T_c-T$, a crossover to the linear dependence on the temperature takes
place in the region $T_c-T\ll T_c$, for sufficiently small $g$.

Some of the CPRs $\tilde{j}(\chi)$ are shown in
Figs.~\ref{anharmon2ab}\,a, b. Except for the first curve in
Fig.~\ref{anharmon2ab}\,b, the strongly anharmonic CPRs in
junctions with large Josephson couplings are displayed. As seen in
Fig.~\ref{anharmon2ab}\,a, the heights of the anharmonic peaks
diminish considerably and the peak positions change weakly, when the
interface pair breaking goes up. Although the anharmonicity can be
well pronounced even in the presence of quite a large pair breaking.
This concerns, in particular, the curve 1 in
Fig.~\ref{anharmon2ab}\,a, which is identical to the curve 3 in
Fig.~\ref{anharmon2ab}\,b shown there in a different scale.
Eq.~\eqref{bcss1012p} describes the CPRs almost perfectly and the
corresponding dashed curves can be distinguished from the exact solid
ones only near the high peak of curve 5 in Fig.~\ref{anharmon2ab}\,a.
All curves in Fig.~\ref{anharmon2ab}\,b are also well approximated by
a simple formula \eqref{cpr3} with deviations (not shown) approaching
only several percent. However, in contrast to \eqref{bcss1012p},
\eqref{cpr3} does not apply to describing upper three curves in
Fig.~\ref{anharmon2ab}\,a. The CPR similar to \eqref{cpr3} was
found earlier within the microscopic description of the dirty
$s$-wave junctions with metallic interlayers \cite{Ivanov1981}.
The strong pair breaking can take place in those junctions, if the
interlayer conductivity considerably exceeds the normal conductivity
of the superconducting metal.

In conclusion, the paper reveals the qualitative features and
develops the quantitative description of the anharmonic Josephson
current near $T_c$. The interface pair breaking as well as
the current depairing and the Josephson coupling-induced pair
breaking have been taken into account and shown to play an important
part in forming the CPR. The results obtained, in particular, concern
the junctions involving $d$-wave superconductors and/or magnetic
interlayers.
\begin{acknowledgments}
The support of RFBR grant 11-02-00398 is acknowledged.
\end{acknowledgments}

\providecommand{\noopsort}[1]{}\providecommand{\singleletter}[1]{#1}%

\end{document}